\documentclass[]{article}

\usepackage{amsmath,amssymb,amsfonts}
\usepackage{graphicx}
\usepackage{color}

\usepackage{authblk}
\usepackage{amsthm}

\usepackage{tikz}
\usetikzlibrary{decorations.pathreplacing}

\usepackage[english]{babel}
\usepackage[utf8]{inputenc}

\usepackage{cite}

\usepackage{appendix}
\usepackage{url}
\usepackage[most]{tcolorbox}

\usepackage[a4paper,margin=1in]{geometry}

\newcommand{\nn}{\nonumber}

\def\a{\alpha}

\def\s{\sigma}

\def\t{\tau}

\def\p{\psi}

\def\<{\langle}
\def\>{\rangle}

\def\tu{{\tilde{0}}}

\def\t2{{\tilde{1}}}

\def\ha{{\hat{a}}}

\def\wtd{{\widetilde{d-1}}}

\begin{document}

	\title{Creating highly symmetric qudit heralded entanglement through highly symmetric graphs}
	
	\author[1]{Seungbeom Chin \thanks{sbthesy@skku.edu}}
 \affil{Department of Electrical and Computer Engineering, Sungkyunkwan University, Suwon 16419, Korea}

\maketitle

\begin{abstract}

Recent attention has turned to exploring quantum information within larger Hilbert spaces by utilizing qudits, which offer increased information capacity and potential for robust quantum communications. While the efficient generation of multipartite qudit entanglement is crucial for studying quantum correlations in high-dimensional Hilbert spaces, the increased dimension makes the circuit design challanging, especially when the entanglement is generated by heralding detections. In this work, we demonstrate that the graph picture of linear quantum networks (LQG picture) can provide a simplified method to generate qudit multipartite heralded entanglement of high symmetries. The LQG picture enables the reduction of circuit complexity by directly imposing the state symmetry onto the circuit structure. Leveraging this insight, we propose heralded schemes for generating $N$-partite $N$-level anti-symmetric (singlet) and  symmetric (Dicke) states. Our study shed light on the optimal circuit design of high-dimensional entanglement with a systematic graphical strategy. 
\end{abstract}

\section{Introduction} \label{intro}

The quantum information science has significantly impacted contemporary technological advancement. Extensive research has been conducted in this field, yielding remarkable outcomes in quantum  computing~\cite{divincenzo1995quantum,knill2001scheme,briegel2009measurement}, communication~\cite{bouwmeester1997experimental,gisin2007quantum}, and simulation~\cite{georgescu2014quantum}. The fundamental unit of quantum information is a qubit, which is constructed with a two-level quantum system and serves as the quantum equivalent of the classical bit. It presents intriguing potential to explore quantum information within larger Hilbert spaces, which include a larger number of qubits or higher-level quantum systems that correspond to  \emph{qudits.} Quantum information processing based on qudits recently draw attention by its increased capacity compared to qubits and potential to execute more robust and secure quantum communications~\cite{cozzolino2019high,wang2020qudits}. For the study of high-dimensinal Hilbert space, it is essential to find efficient schemes for generating multipartite qudit entanglement with realistic quantum particles. 

One promising method for tackling this task involves leveraging the indistinguishability of quantum particles. Several studies have proposed both theoretical and experimental schemes for generating entanglement, utilizing the identity of quantum  particles with postselection (see, e.g.,~\cite{tichy2013entanglement,krenn2017entanglement,chin2019entanglement,barros2020entangling,blasiak2019entangling, lee2022entangling}).
However, in postselected 
schemes, the generation of the target state is confirmed only after the detection of particles, which makes the resource less suitable for consecutive quantum gate operations. This limitation serves as significant motivation for investigating heralded schemes~\cite{barz2010heralded,papp2009characterization,zeuner2018integrated,li2021heralded,le2021heralded,chin2022graph,chin2023graphs,chin2023linear},  which integrate ancillary particles and modes to indicate the successful generation of target states.
This approach enables the selection of experimental runs that yield the desired target states without directly measuring them. Consequently, the entanglement produced through heralded schemes becomes more handy resource that is not destructed by detectors. Heralded entangles states are also considered useful for loophole-free Bell violation tests by their robust structures against lossy channels~\cite{zhao2019higher,weston2018heralded}.

Nevertheless, heralded schemes typically require much more particles and modes for heralding resources, rendering them more intricate to design compared to postselected schemes.
To overcome this limitation of heralded schemes, a graph-based approach for generating heralded entanglement is introduced Ref.~\cite{chin2022graph,chin2023graphs,chin2023linear}, which establish a framework wherein many-particle systems, including  boson annihilation operators, are mapped to bipartite graphs (bigraphs). This is an extension of the graph picture introduced in Ref.~\cite{chin2021graph}. Since there are other several graphical methods to understand quantum information processing~\cite{van2020zx,biamonte2019lectures,hein2006entanglement,krenn2017quantum}, we term our method the \emph{linear quantum graph (LQG) picture} to avoid confusion. 

In this work, we exploit the LQG picture to find intriguing and valuable qudit multipartite entangled states with heralding. The circuit complexity increases substantially when we increase the dimension of Hilbert space and simultaneously choose heralding instead of postselection. However, when the state has a highly symmetric structure, the LQG picture provides a powerful insight to reduce the design complexity by directly imposing the state symmetry to the circuit itself. One of such examples is the qudit GHZ state generating scheme proposed in Ref.~\cite{chin2022graph}, which shares the same cyclic symmetry with the target state. We suggest heralded schemes for generating $N$-partite $N$-level symmetric and anti-symmetric states based on the property.

The $N$-partite $N$-level anti-symmetric (singlet) state $|\mathcal{S}_N\>$ is written as
   \begin{align}
        |\mathcal{S}_N\> \equiv \frac{1}{\sqrt{N!}}\sum_{\sigma\in S_N} sgn(\sigma) |\s(0),\s(1),\cdots, \s(N-1) \>
    \end{align} ($S_N$ is the permutation group and $sgn(\sigma)$ is the signiture of $\sigma$), 
which is known to be useful resource for, e.g., the problems of 
$N$-strangers, secret sharing, liar detection~\cite{cabello2002n}, and  byzantine agreement~\cite{fitzi2001quantum}. 

Among several types of qudit Dicke states (see, e.g., Ref.~\cite{nepomechie2023qudit} for the definition), we focus on the following state: 
   \begin{align}
       |D^N(\underbrace{1,1,\cdots, 1}_{N})\> \equiv \frac{1}{\sqrt{N!}}\sum_{\sigma\in S_N} |\s(0),\s(1),\cdots, \s(N-1) \>,
    \end{align} which is one of the states that has the largest minimal decomposition entropy~\cite{enriquez2016maximally}  and useful for quantum communication and metrology~\cite{li2021verification}. 

So far as we know, there have been no scheme for the heralded generations of the $N$-partite $N$-level totally symmetric and anti-symmetric states. For the case of $|\mathcal{S}_N\>$, there are a few attempts to generate the state with quantum nondemolition (QND) measurements~\cite{toth2010generation,piccolini2023robust}, however they are restricted by  the low efficiency of  QND detectors. On the other hand, our scheme consist of linear transformation operators and conventional number-resolution detector, hence more feasible.

\section{LQG picture of $d$-level  Qudit entanglement generation with boson subtractions}
	

	\begin{figure}
		\centering
		\includegraphics[width=8cm]{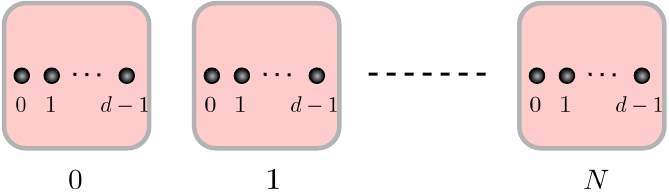}
		\caption{The initial boson distribution of $dN$ bosons in the qudit sculpting protocol. Each spatial mode from 1 to $N$ has $d$ bosons, whose internal states are  mutually orthogonal with each other, i.e. $|0\>, |1\>,\cdots, |d-1\> $. } 
		\label{fig_initial_qudit}
	\end{figure}

 \begin{table}[h]
\centering 
			\begin{tabular}{|l|l|l|}
				\hline
				\textbf{Boson systems with a sculpting operator} & \textbf{Bipartite graph $G_b =(U\cup V, E)$}  \\
				\hline\hline 
				Spatial modes & Labelled vertices $\in$ $U$  \\ \hline 
				$\hat{A}^{(l)}$  ($l \in \{1,2,\cdots, N\}$) & Unlabelled vertices $\in$  $V$ \\ \hline 
				Spatial distributions of $\hat{A}^{(l)}$  & Edges  $\in$ $E$ \\ \hline 
				Probability amplitude $\a_j^{(l)} $ & Edge weight $\a_j^{(l)} $ \\ \hline 
				Internal state $\p_j^{(l)} $ & Edge weight $\p_j^{(l)}$  \\
				\hline 
			\end{tabular}
   \caption{Correspondence relations of a sculpting operator to a sculpting bigraph}
   			\label{dict}
		\end{table} 

We design our heralded schemes following the \emph{$d$-level boson sculpting protocol}, which is presented in Ref.~\cite{chin2022graph} by generalizing the the 2-level boson sculpting protocol~\cite{karczewski2019sculpting}.

In our system, bosons are in $j$th spatial mode ($j\in \{1,2,\cdots, N\}$) with a $d$-dimensional internal degree of freedom $s$ $(\in \{0,\cdots, d-1\})$, hence we denote the boson creation (annihilation) operators as $\ha_{j,s}^\dagger$ ($\ha_{j,s}$). To create $N$-partite $d$-level entanglement, we first allocate $dN$ bosons into $N$ spatial modes evenly as in Fig.~\ref{fig_initial_qudit}, which is represented in the operational form by
	\begin{align}\label{initial}
		|Sym_{N,d}\> \equiv \prod_{j=1}^N(\ha_{j,0}\ha_{j,1}\cdots \ha_{j,d})|vac\>.
	\end{align}
	
Then we subtract $(d-1)N$ bosons from the $N$-partite system with a sculpting operator $\hat{A}_N= \prod_{l=1}^{(d-1)N}\hat{A}^{(l)}$
so that it subtracts $(d-1)$ particles per mode. Hence the final state is given by
\begin{align}
  |\Psi\>_{fin}= \hat{A}_N|Sym_{N,d}\>
\end{align}

Then each particle remaining in each mode constitutes qudit information. A suitable sculpting operator generates a genuine multipartite entangled state.

The LQG picture of the sculpting operator is suggested in Ref.~\cite{chin2022graph}, which is displayed in Table~\ref{dict}. For a clear visibility of graphs, we employ a simplified notation for edge weights. More specifically, in this work we consider only cases when all the absolute values of probability amplitude weights are equal  among those which are attached to the same dot. Hence, we omit the probability amplitude weights when the phase of the probabiltiy amplitude is $0$, and denote phase (not amplitude) otherwise. In addition, we work in a computational basis $\{0,1,\cdots, d-1 \}$ and its Fourier-transformed basis $\{\tilde{0},\tilde{1},\cdots, \wtd \}$ in our approach, which can be replaced with edge colors. We set $\tilde{0} =$Red and $\wtd 
=$Blue in this work. 

Now we provide a crucial property of  bigraphs that is essential to design qudit multipartite entanglement: For a bigraph, if edges are attached to all the circles as 
    \begin{align}\label{edm_special}
        \includegraphics[width=2cm]{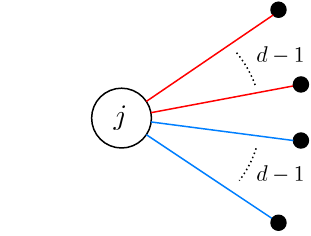}, 
    \end{align} then the final state is given by the summation of the $(d-1)$-to-one matchings between the circles and dots of the given bigraph, which correspond to the terms of the final state.   
This property is directly shown with the following identity
\begin{align}
(\ha_{\tu})^l(\ha_{\widetilde{d-1}})^{d-1-l}\prod_{s=0}^{d-1}\ha^\dagger_{s} |vac\>=(-1)^{d-1-l}\frac{l!(d-1-l)!}{\sqrt{d}^{d-2}}
\ha^\dagger_{\widetilde{d-1-l}} |vac\>~~~~(l\in \{ 0,1,\cdots ,d-1 \})
\end{align} and its corresponding graph identity. 

The property is essential to find entanglement generation schemes, because our sculpting bigraphs have $N$ circles and $(d-1)N$ dots, whose structures are set so that only $(d-1)$-to-one matchings contribute to the final state in the $d$-level sculpting protocol. This is a generalization of EPM bigraphs, in which 
only perfect matchings contribute to the final state by the two-level sculpting protocol~\cite{chin2021graph}.

An example that satisfies the above restriction~\eqref{edm_special} is already given in Ref.~\cite{chin2022graph},
\begin{align}
\begin{gathered}
       \includegraphics[width=4cm]{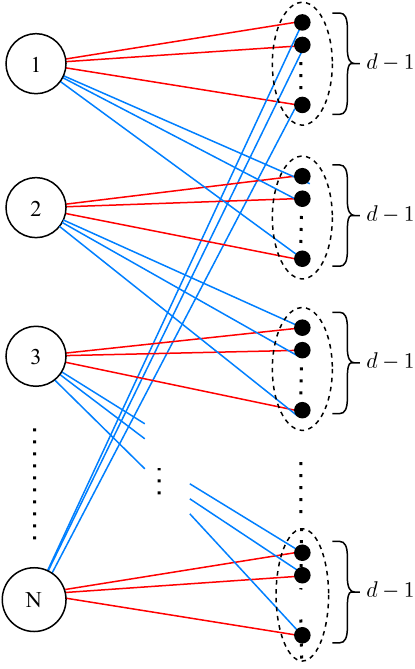} 
\end{gathered}
\end{align} which corresponds to the sculpting scheme to generate $N$-partite qudit  GHZ state (note that the cyclic symmetry of GHZ state is reflected in the graph).

\section{Sculpting schemes through the graph symmetries}\label{sculpting_scheme}

In this section, we present sculpting bigraphs of $N$ circles and $(N-1)N$ dots that generate $|\mathcal{S}_N\>$ and $|D^N(1,1,\cdots, 1)\>$.
We will impose the total (anti-) symmetry under the circle label exchanges to those graphs, which directly corresponds to the total   (anti-) symmetry under the spatial modes of the final target states. To see this intuitively, note that  
 the total (anti-) symmetry is decided by $\frac{N(N-1)}{2}$-pairs of relations. We can impose such relations to graphs with $N(N-1)$ edges that connect dots and circles, hence two edges attached to two dots are combined to produce such (anti-) symmetric property to graphs.

\subsection{Totally anti-symmetric sculpting bigraph}

To make the sculpting bigraph totally anti-symmetric, 
we use the following subgraph:
\begin{align}\label{anti_symm_element}
\begin{gathered}
   \includegraphics[width=2.5cm]{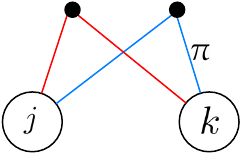}
\end{gathered}
 = (\ha^\dagger_{j,\tilde{0}} + \ha^\dagger_{k,\tilde{0}})(\ha^\dagger_{j,\widetilde{N-1}}-\ha^\dagger_{k,\widetilde{N-1}})~~~~~(j <k).
\end{align} 
This subgraph is anti-symmetric under the exchange of modes $j$ and $k$. Since $j,k \in \{1,2,\cdots, N\}$, there are $\frac{N(N-1)}{2}$ subgraphs that contruct the totally anti-symmetric sculpting bigraphs. This sculpting bigraph is equivalent to the sculpting operator
\begin{align}
    \hat{A}^{\mathcal{S}}_{N} = \prod_{j<k =0}^{N-1} (\ha^\dagger_{j,\tilde{0}} + \ha^\dagger_{k,\tilde{0}})(\ha^\dagger_{j,\widetilde{N-1}}-\ha^\dagger_{k,\widetilde{N-1}}).
\end{align}
The sculpting bigraphs for $N=3,4,$ and 5  are drawn as
\begin{align}\label{anti_bigraph}
      \includegraphics[width=.75\textwidth]{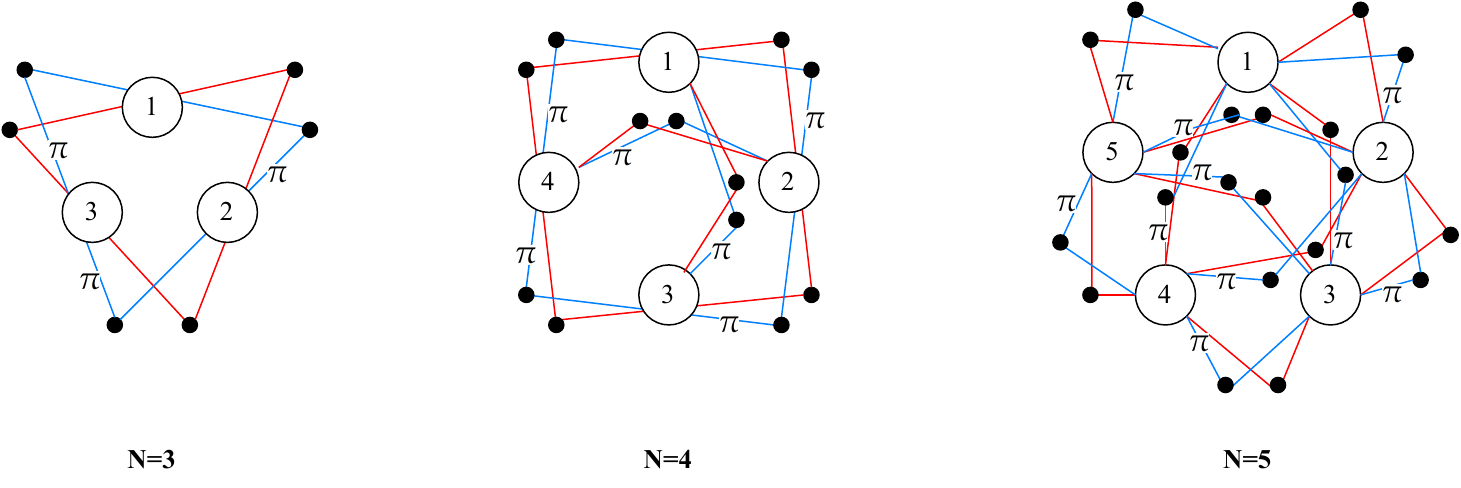} 
\end{align}
It is direct to see that this bigraph is anti-symmetric under any exchange between two circles by the anti-symmetric property of the subgraph \eqref{anti_symm_element}, which also can be directly checked in the operator side. Since the initial state $|Sym_{N,d}\>$ is totally symmetric, we can see that the final state is totally anti-symmetric.

\subsection{$N$-partite $N$-level Dicke state $|D^N(\underbrace{1,1,\cdots, 1}_{N})\>$ sculpting bigraph}

For the generation of the totally symmetric state $|D^N(\underbrace{1,1,\cdots, 1}_{N})\>$, we use a slightly different subgraph
\begin{align}
\begin{gathered}
     \includegraphics[width=2.5cm]{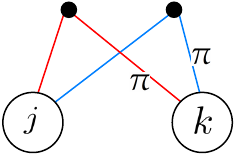}    
\end{gathered}
      = (\ha^\dagger_{j,\tilde{0}} -\ha^\dagger_{k,\tilde{0}})(\ha^\dagger_{j,\widetilde{N-1}}-\ha^\dagger_{k,\widetilde{N-1}}).
\end{align}
$\frac{N(N-1)}{2}$ of the above subgraphs constitute the Dicke sculpting bigraph, which is equivalent to the sculpting operator
\begin{align}
    \hat{A}^{D_1} = \prod_{j<k =0}^{N-1} (\ha^\dagger_{j,\tilde{0}} -\ha^\dagger_{k,\tilde{0}})(\ha^\dagger_{j,\widetilde{N-1}}-\ha^\dagger_{k,\widetilde{N-1}}).
\end{align}

The sculpting bigraphs for $N=3,4,$ and 5 are drawn as
\begin{align}\label{symm_bigraph}
\includegraphics[width=.8\textwidth]{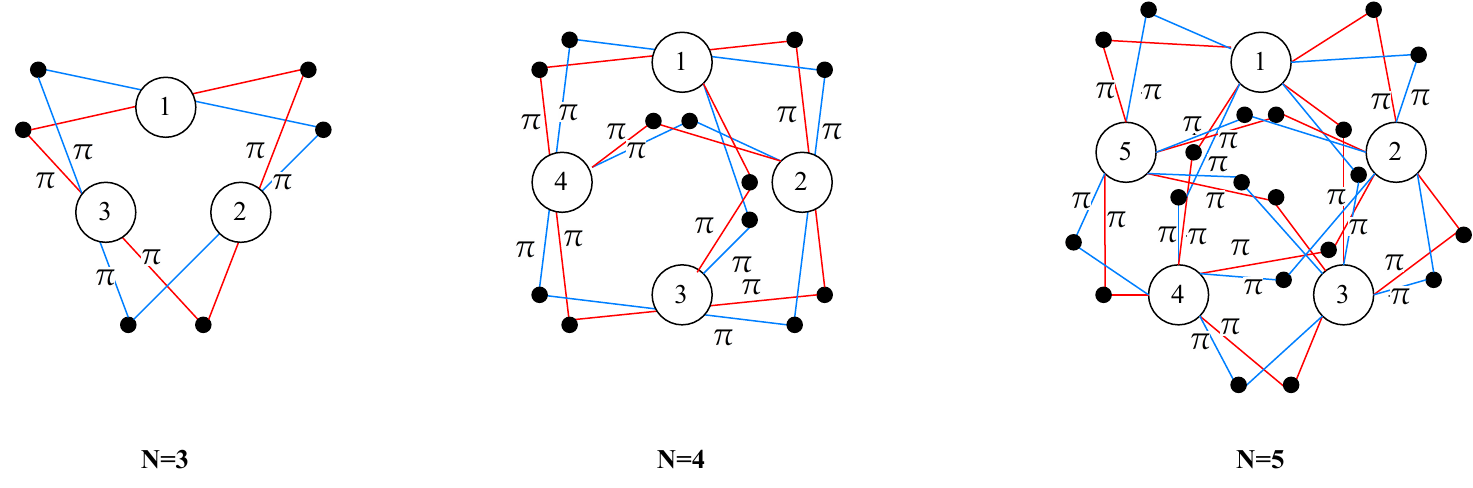} \nn 
\end{align}

To see that this type of bigraph generates $|D^N(\underbrace{1,1,\cdots, 1}_{N})\>$, we need to devide two subgraphs that are just made of the same color of edges, which correspond to the operators  
$\prod_{j<k =0}^{N-1}(\ha_{j,0} -\ha_{k,0})$ and $\prod_{j<k =0}^{N-1}(\ha_{j,\widetilde{d-1}} -\ha_{k,\widetilde{d-1}})$. These operators are totally anti-symmetric under the exchange of spatial modes respectively, which combine to make the overall operator $A_{N}^{D_1}$ symmetric. This implies that in each operator terms of the same order of annihilation operators cancel with each other to conserve the anti-symmetry. Hence, when the operator is applied to the initial state, the final state consists of terms whose internal states are all different with each other, which is  $|D^N(\underbrace{1,1,\cdots, 1}_{N})\>$.

In addition, it is worth mentioning that 
\begin{align}
\begin{gathered}
     \includegraphics[width=2.5cm]{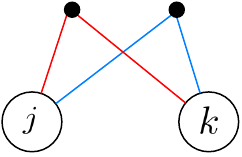}    
\end{gathered}
      =  (\ha^\dagger_{j,\tilde{0}} + \ha^\dagger_{k,\tilde{0}})(\ha^\dagger_{j,\widetilde{N-1}}+\ha^\dagger_{k,\widetilde{N-1}})
\end{align} also generates another totally symmetric state. However, since  $\prod_{j<k =0}^{N-1}(\ha_{j,0} +\ha_{k,0})$ and $\prod_{j<k =0}^{N-1}(\ha^\dagger_{j,\widetilde{N-1}}+\ha^\dagger_{k,\widetilde{N-1}})$ are both symmetric, the final state is made of more terms. For example, for $N=3$, the final state is given by
\begin{align}
&\prod_{j<k=0}^{2}(\ha^\dagger_{j,\tilde{0}} +\ha^\dagger_{k,\tilde{0}})(\ha^\dagger_{j,\widetilde{2}}+\ha^\dagger_{k,\widetilde{2}})|Sym_{3,3}\>\nn \\
&=
|012\> + |120\>+|201\> +|210\>+|102\>+|021\> +|111\> \nn \\
   &= |D^3(1,1,1))\> +|111\>.
\end{align}
Note that this final state is similar to the qutrit tripartite Dicke state $|D_3^3\>$ defined in Ref.~\cite{laskowski2014noise},
\begin{align}
|D_3^3\> = |012\> + |120\>+|201\> +|210\>+|102\>+|021\> +2|111\> .  
\end{align}




\section{Linear optical schemes with heralding detectors}

We design linear optical circuits with heralding detectors based on the sculpting operators introduced in Section~\ref{sculpting_scheme}. A set of translation rules from qubit sculpting bigraphs to linear optical circuits are presented in Ref.~\cite{chin2023graphs}, which can be applied to qudit cases by straightforward generalization. Here we display the $N=3$ examples, which is directly generalized to larger $N$ cases. 

To design linear optical circuits that proceed in time from left to right, we re-array the $N=3$ sculpting bigraphs in~\eqref{anti_bigraph} and \eqref{symm_bigraph} for the $N=3$ anti-symmetric and symmetric states as
\begin{align}\label{n3_vertical}
    \includegraphics[width=3cm]{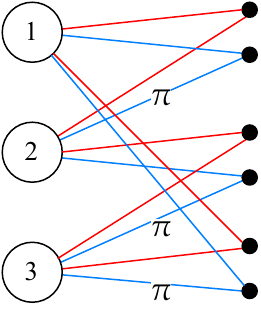},~~~~~~~~~~~
      \includegraphics[width=3cm]{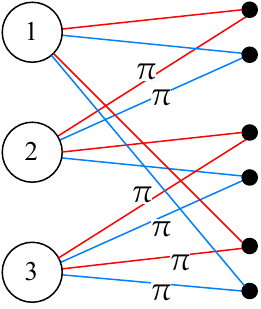}.
\end{align}

Considering each sculpting operator as a heralding detector, the $N=d=3$ totally anti-symmetric state is generated by the following optical circuit with triple-rail encoding:

\begin{align}
\begin{gathered}
    \includegraphics[width=1\textwidth]{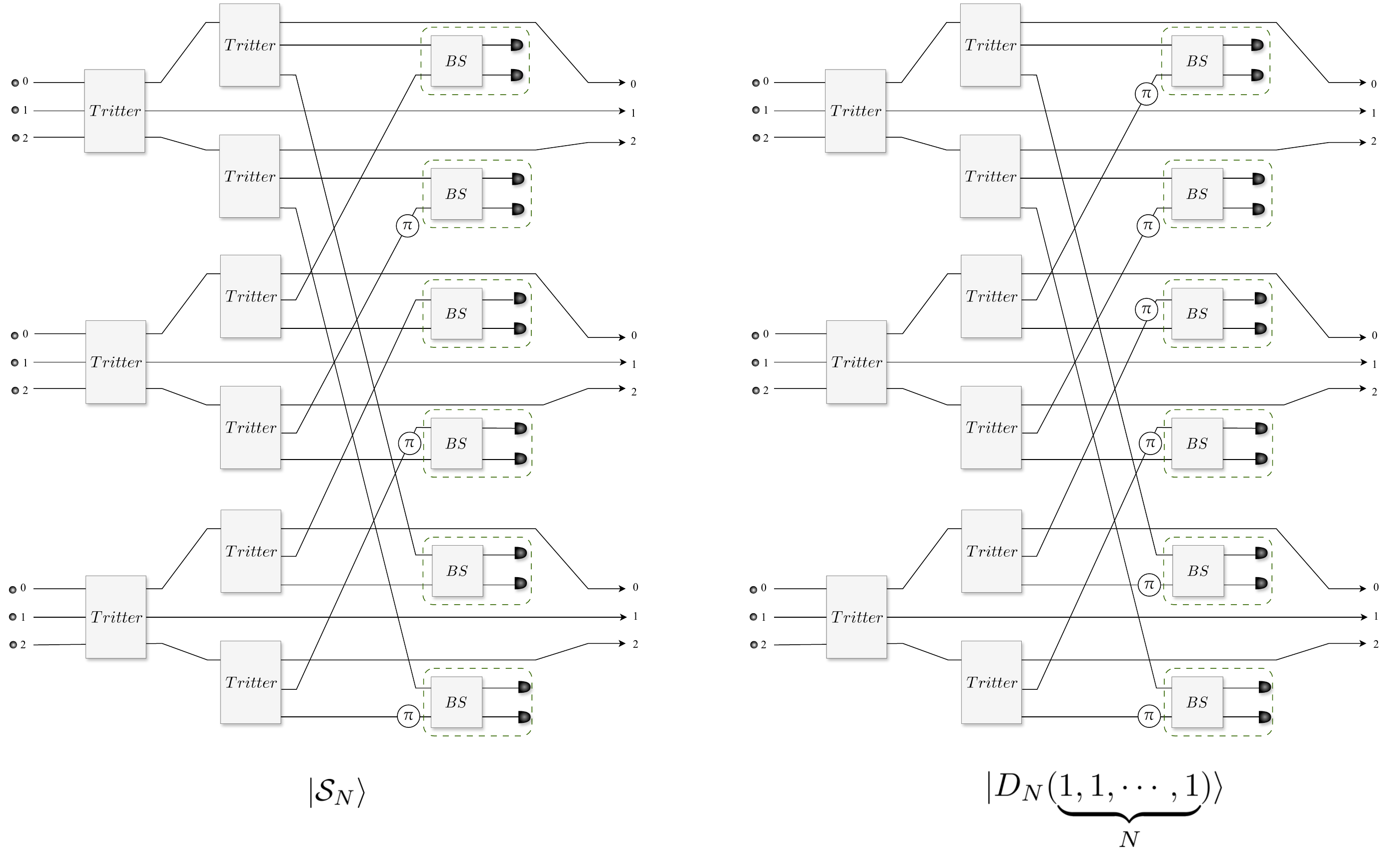}
    \end{gathered}
    \nn 
\end{align} where
\begin{align}
  &  \begin{gathered}
        \includegraphics[width=2.5cm]{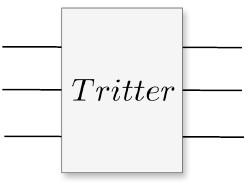}
    \end{gathered}~:~j\to\tilde{j}~\{0,1,2\},~~   \begin{gathered}
        \includegraphics[width=1.6cm]{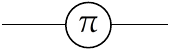}
    \end{gathered}~:~\textrm{Phase rotation by }\pi,\nn \\
   &~~~~\begin{gathered}
        \includegraphics[width=2cm]{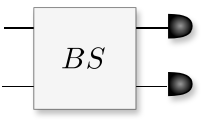}
            \end{gathered}~:~\textrm{Postselection of one-particle states after the beam splitter }BS
\end{align} Note that the operations in the dashed boxes of the above circuits play the role of subtraction operators. Wires are connected to those boxes following the structure of sculpting bigraphs~\eqref{n3_vertical}.  
The success probability is $\frac{2\sqrt{6}}{3^8}$ with feed-forward at the detectors after BSs.

The $N$-partite $N$-level generalization of the above schemes is straighforward by replacing tritters with $N$-partite ports, whose success probability is given by $\frac{N!\sqrt{N!}}{(N^N)^N}$.

\section{Discussions}

In this study, we have proposed heralded schemes that generate $N$-partite $N$-level symmetric and anti-symmetric states with the LQG picture. 
We expect our current results serve as a foundation upon which to build interesting systematic and optimized strategies for the generation of qudit multipartite entangled states. 
Our graphical solutions can be considered a special group of the qudit generalization of EPM bigraphs introduced in Ref.~\cite{chin2022graph}, which has convenient properties that can directly correspond to the heralded generation of $qubit$ multipartite entanglement. A more rigorous formalism for the $qudit$ generalization of EPM bigraphs including more ancillary particles will provide abundance possibility to find more complicated qudit entangled states with less symmetries, such as absolutely maximally entangled states and qudit graph states.

\section*{Acknowledgements}

This research was supported by
National Research Foundation of Korea (NRF, RS-2023-00245747) and the quantum computing technology development program of the National Research Foundation of Korea (NRF) funded by the Korean government (Ministry of Science and ICT, MSIT, No.2021M3H3A103657313).

\bibliographystyle{unsrt}
\bibliography{anti_symm}

\end{document}